\documentclass[11pt,letter]{article}
\usepackage[intlimits,centertags]{amsmath}
\usepackage{amssymb,amsfonts}
\usepackage{bbm}
\usepackage{mathrsfs}
\usepackage{cite}
\usepackage[scale={0.75,0.75}]{geometry}
\usepackage{multirow}
\usepackage{relsize}
\usepackage[subrefformat=parens,labelformat=parens]{subfig}
\usepackage{color}
\usepackage{caption}
\usepackage[usenames,dvipsnames]{xcolor}
\usepackage{graphics,graphicx}
\newlength{\textlength}
\newlength{\barlength}

\allowdisplaybreaks[2]
\numberwithin{equation}{section}

\begin{document}

%\title{\vspace{2cm}\textbf{No-scale D-term inflation with stabilized moduli}}
%\author{Wilfried Buchm\"{u}ller, Valerie Domcke, Clemens Wieck\\[2mm]
%    {\normalsize\itshape Deutsches Elektronen-Synchrotron}\\{\normalsize\itshape Hamburg, Germany}\\ 
%    {\small\ttfamily wilfried.buchmueller, valerie.domcke, clemens.wieck @desy.de}}
%  \date{\today}
%\maketitle

\noindent DESY 13-157 \newline
September 2013

\vspace{3.5cm} \begin{center}
\LARGE{\textbf{No-scale D-term inflation with stabilized moduli}} 

\vspace{1cm}

\large{Wilfried Buchm\"{u}ller, Valerie Domcke, Clemens Wieck}\\[2mm]
    {\normalsize\itshape Deutsches Elektronen-Synchrotron}\\{\normalsize\itshape Hamburg, Germany}\\ 
    {\small\ttfamily wilfried.buchmueller, valerie.domcke, clemens.wieck @desy.de}

%\vspace{5mm}

%\today

\end{center}

\vspace{1cm}

\begin{abstract}

\noindent 
We study the consistency of hybrid inflation and moduli stabilization, using the Kallosh--Linde model as an example for the latter. We find that F-term hybrid inflation is not viable since inflationary trajectories are destabilized by tachyonic modes. On the other hand, D-term hybrid inflation is naturally compatible with moduli stabilization due to the absence of a large superpotential term during the inflationary phase. Our model turns out to be equivalent to superconformal D-term inflation and it therefore successfully accounts for the CMB data in the large-field regime. Supersymmetry breaking can be incorporated via an O'Raifeartaigh model. For GUT-scale inflation one obtains stringent bounds on the gravitino mass. A rough estimate yields $10^{5}\,\text{GeV}\lesssim m_{3/2} \lesssim 10^{10}\,\text{GeV}$, contrary to naive expectation.
\end{abstract}
\thispagestyle{empty}
\clearpage

%
%%
%%%
%%%%
%%%%%%%%%%%%%%%%%%%%%%%%%%%%%%%%%%%%%%%%%%%%%%%
%%%%%%%%%%%%%%%%%%%%%%%%%%%%%%%%%%%%%%%%%%%%%%%
%%%%%%%%%%%%%%				 %%%%%%%%%%%%%%%%%%%%%%%
%%%%%%%%%%%%%	%         Section 1        %%%%%%%%%%%%%%%%%%%%%%%
%%%%%%%%%%%%%%				  %%%%%%%%%%%%%%%%%%%%%%
%%%%%%%%%%%%%%%%%%%%%%%%%%%%%%%%%%%%%%%%%%%%%%%
%%%%%%%%%%%%%%%%%%%%%%%%%%%%%%%%%%%%%%%%%%%%%%%

\section{Introduction} 
\label{sec:Introduction}

Hybrid inflation \cite{Linde:1993cn} is an attractive mechanism for generating the cosmological density perturbations. It is naturally realized in the framework of grand unified theories (GUTs) and string theory, as F-term \cite{Copeland:1994vg,Dvali:1994ms} or D-term inflation \cite{Binetruy:1996xj,Halyo:1996pp} where the GUT scale emerges through the Fayet-Iliopoulos (FI) term of an anomalous $U(1)$ symmetry. However, the embedding of hybrid inflation in a UV-complete theory, which has significant effects on GUT-scale inflation, remains an open question.

The probably best-motivated UV-complete theory for this embedding is string theory. In this framework, six dimensions have to be compactified on a Calabi--Yau manifold to obtain a four-dimensional effective theory with $\mathcal N = 1$ supersymmetry. In the classical perturbative four-dimensional theory massless scalar fields, so-called moduli, arise as remnants of the internal manifold. The stabilization of these moduli has been a widely discussed subject for many years.
In type IIB string compactifications on Calabi--Yau manifolds with D-branes and fluxes, it has been shown that all complex structure moduli and the axio-dilaton can be stabilized by fluxes \cite{Giddings:2001yu}. K\"ahler moduli, on the other hand, can be stabilized by non-perturbative contributions to the superpotential, such as gaugino condensates on stacks of D-branes \cite{Kachru:2003aw}. The latter have been used in a model by Kallosh and Linde (KL) \cite{Kallosh:2004yh}, where a single K\"ahler modulus is stabilized in a racetrack potential with vanishing vacuum energy in a local minimum. This setup has the appealing feature of scale separation between the Hubble scale $H_\text{inf}$ during inflation and the gravitino mass, which can be very small compared to $H_\text{inf}$.

In this paper, we study the effects of stabilizing the K\"ahler modulus in such a racetrack potential on the dynamics of hybrid inflation. As was pointed out in~\cite{Brax:2006ay}, even a tiny displacement of the modulus field due to its gravitational coupling to the inflaton field can be fatal for a potential inflationary trajectory, as can be seen explicitly by integrating out the modulus field. Our work is related to earlier attempts of combining hybrid inflation and moduli stabilization in F-term~\cite{Brax:2006ay, Davis:2008sa, Mooij:2010cs} and D-term inflation~\cite{Brax:2006yq} as well as in chaotic inflation~\cite{Davis:2008fv}. Here, we use a specific form of K\"ahler potential, motivated by the no-scale K\"ahler potential of the modulus field and an approximate superconformal symmetry. Similar to~\cite{Brax:2006ay, Davis:2008sa}, we find that F-term hybrid inflation is spoiled by corrections induced by the modulus sector. In particular, whenever one direction of the complex inflaton is flat, the other one is tachyonic. However, we find that D-term hybrid inflation can be successfully combined with moduli stabilization.

The resulting no-scale D-term inflation model has a number of interesting features. Along the inflationary trajectory it is actually equivalent to the superconformal D-term inflation model proposed in~\cite{Buchmuller:2012ex}. As shown in \cite{Buchmuller:2013zfa}, in the large field regime it asymptotically yields the Starobinsky model~\cite{Starobinsky:1980te}, which agrees remarkably well with the recently released Planck data~\cite{Ade:2013uln}. Supersymmetry breaking can be accomplished by adding a quantum corrected O'Raifeartaigh model \cite{Kallosh:2006dv} without spoiling moduli stabilization or inflation. For GUT-scale inflation one obtains stringent bounds on the gravitino mass.

\medskip

This paper is organized as follows. Our scheme of racetrack moduli stabilization and its coupling to F-term hybrid inflation is discussed in Section~\ref{sec:2}. Turning to D-term hybrid inflation in Section~\ref{sec:3}, we calculate all relevant corrections to the inflationary dynamics arising from moduli stabilization, summarize the inflationary predictions, and discuss supersymmetry breaking in this context. We conclude in Section~\ref{sec:Conclusion}.

%
%%
%%%
%%%%
%%%%%%%%%%%%%%%%%%%%%%%%%%%%%%%%%%%%%%%%%%%%%%%
%%%%%%%%%%%%%%%%%%%%%%%%%%%%%%%%%%%%%%%%%%%%%%%
%%%%%%%%%%%%%%				 %%%%%%%%%%%%%%%%%%%%%%%
%%%%%%%%%%%%%	%         Section 2        %%%%%%%%%%%%%%%%%%%%%%%
%%%%%%%%%%%%%%				  %%%%%%%%%%%%%%%%%%%%%%
%%%%%%%%%%%%%%%%%%%%%%%%%%%%%%%%%%%%%%%%%%%%%%%
%%%%%%%%%%%%%%%%%%%%%%%%%%%%%%%%%%%%%%%%%%%%%%%

\section{F-term hybrid inflation}
\label{sec:2}

In its simplest form, the superpotential of F-term hybrid inflation in terms of the chiral superfields $S$, $\phi_+$, and $\phi_-$ can be written as \cite{Copeland:1994vg}
\begin{align}\label{eq:fhybsuperpot}
W_\text{HI} = \lambda S \left( \phi_+ \phi_- - v^2 \right)\,.
\end{align}
In this setup, $S$ contains the inflaton and $\phi_\pm$ are the so-called waterfall fields, carrying charge~$\pm q$ under some local $U(1)$ symmetry, which are responsible for ending inflation. Moreover, $v$ is of the same order as the GUT scale and the coupling $\lambda$ is chosen to be real.

\medskip

The slow-roll potential for the inflaton is typically generated by supergravity interactions and the one-loop Coleman--Weinberg potential. At a critical field value $S_\text{c} = v$ the waterfall fields obtain a tachyonic mass and inflation ends with spontaneous symmetry breaking of the $U(1)$ symmetry. For a more detailed account of the dynamics and phenomenology of F-term hybrid inflation in supergravity, see e.g. \cite{Buchmuller:2000zm,Linde:1997sj}.

%
%%
%%%
%%%%
%%%%%%%%%%%%%%%%%%%%%%%%%%%%%%%%%%%%%%%%%%%%%%%
%%%%%%%%%%%%%%%%%%%%%%%%%%%%%%%%%%%%%%%%%%%%%%%
\subsection{KL moduli stabilization}

When hybrid inflation is embedded in a higher-dimensional theory, the question of moduli stabilization has to be addressed. For simplicity, we consider a scenario in which the overall volume of the compactified dimensions is parameterized by a single K\"ahler modulus $\rho=\sigma + i \beta$. This case is well understood in type IIB string theory. In particular, it is assumed that the dilaton and all complex structure moduli have been stabilized by fluxes \cite{Giddings:2001yu} and only one K\"ahler modulus remains massless. This K\"ahler modulus can be stabilized by non-perturbative contributions to the superpotential \cite{Kachru:2003aw,Kallosh:2004yh} in combination with a no-scale K\"ahler potential,
\begin{align}\label{eq:KLKahlerpot}
K = -3 \ln{(\rho+\bar \rho)}\,.
\end{align}
In case of two non-perturbative terms, the superpotential reads
\begin{align}\label{eq:KLsuperpot}
W_\text{KL} = W_0 + A e^{-a \rho} + B e^{-b \rho}\,.
\end{align}
Here, $W_0$, $A$, and $B$ are determined by fluxes, and the non-perturbative terms in eq.~\eqref{eq:KLsuperpot} are generated by gaugino condensates on stacks of D-branes. The parameters $a$ and $b$ are given by $\frac{2 \pi}{N_i}$, $i \in \{a,b\}$, where $N_i$ are the ranks of the condensed gauge groups.

\medskip

In the model of Kallosh and Linde \cite{Kallosh:2004yh}, $W_0$ is adjusted to produce a supersymmetric Minkowski vacuum. The minimum of $V$ occurs at $\beta = 0$ and 
\begin{align}
\sigma \equiv \sigma_0= \frac{1}{a-b} \ln{\left|\frac{aA}{bB}\right|}\,. 
\end{align}
This is achieved by choosing 
\begin{align}\label{eq:KLW0}
W_0 = -A \left( \frac{a A}{b B} \right)^{\frac{a}{b-a}} - B \left( \frac{aA}{bB} \right)^{\frac{b}{b-a}}\,,
\end{align}
such that $W_\text{KL}(\sigma_0)=D_\rho W_\text{KL}(\sigma_0)=V(\sigma_0)=0$. In this setup, the modulus is generically very heavy,
\begin{align}
m_\rho^2= \frac29  (a-b) a b A B \ln{\left( \frac{a A}{b B}\right)} \left( \frac{a A}{b B} \right)^{-\frac{a+b}{a-b}}\,,
\end{align}
so that $m_\rho \sim \mathcal OÊ\left( 10^{-3} - 10^{-1} \right)$ in Planck units, for typical parameter values. Note that the compactified dimensions have to be stabilized at large enough volume $\mathcal V = \mathcal O( \sigma_0^{3/2})$ to satisfy both the supergravity approximation and the single-instanton approximation of this analysis. In particular, it is required that $\sigma_0 \gg 1$ and $a \sigma_0, \,b \sigma_0 \gg 1$. In the following, we assume ${\sigma_0 = \mathcal O (10-100)}$ for typical values of the racetrack parameters.

%
%%
%%%
%%%%
%%%%%%%%%%%%%%%%%%%%%%%%%%%%%%%%%%%%%%%%%%%%%%%
%%%%%%%%%%%%%%%%%%%%%%%%%%%%%%%%%%%%%%%%%%%%%%%
\subsection{Effective scalar potential}

Combining the two sectors discussed above to a model with superpotential 
\begin{align}\label{eq:combinedsuperpot}
W = W_\text{KL} + W_\text{HI}\,,
\end{align}
with unspoiled inflation turns out to be a non-trivial task. As pointed out in \cite{Brax:2006ay,Davis:2008sa}, even when the modulus mass is larger than the inflationary Hubble scale, supergravity corrections from the modulus sector generically ruin inflation. During the slow-roll phase the minimum of the modulus potential is slightly shifted, causing the modulus to move by an amount $\delta\rho$ during inflation. The back-reaction of this shift generates a large mass for the inflaton so that $\eta = \mathcal O(1)$. This problem persists when using a no-scale K\"ahler potential with or without a shift symmetry for the inflaton \cite{Davis:2008sa}.

However, this $\eta$-problem can be overcome using a particular K\"ahler potential,
\begin{align}\label{eq:Ftermkahler}
K = -3 \ln{\left[\rho + \bar \rho -\frac13 \left( |S|^2 + |\phi_+|^2 + |\phi_-|^2 \right) -\frac{\chi}{6} \left( S^2+\bar S^2 \right) \right]} \equiv -3 \ln{X} \,,
\end{align}
with $\chi \in \mathbbm R$, which has approximate no-scale form \cite{Cremmer:1983bf} with an $SU(1,3)$ symmetry broken only by the term proportional to $\chi$. As discussed in Section \ref{sec:Starobinsky} this type of K\"ahler potential is also well motivated from the underlying superconformal symmetry of supergravity (see, e.g.~\cite{Ferrara:2010in}). Note that for $\chi=1$ eq.~\eqref{eq:Ftermkahler} reproduces the standard form of a shift symmetric no-scale K\"ahler potential. Using eq.~\eqref{eq:combinedsuperpot} and eq.~\eqref{eq:Ftermkahler} the scalar potential during inflation, i.e., at $S>v$ and $\phi_+ = \phi_-=0$, reads
\begin{align} \nonumber
V &= \frac{1}{X^2}\Bigg\{ \lambda^2 v^4 + \frac13 \left( X + \frac13 |S+\chi \bar S|^2\right)|W_\text{KL}'|^2 \\ & \hspace{1.4cm} - \left[ \left(\overline W - \frac13 \lambda v^2 \left(\bar S + \chi S \right) \right) W_\text{KL}' + \text{c.c.} \right] \Bigg\} \,,
\label{eq:Ftermtreepot}
\end{align}
where primes denote derivatives with respect to $\rho$. To take the shift of the modulus during inflation into account we expand eq.~\eqref{eq:Ftermtreepot} in the displacement $\delta \rho = \rho-\sigma_0$, where $\sigma_0$ denotes the minimum of the pure KL potential, i.e., the minimum after inflation. Thus, we compute the effective potential
\begin{align}\label{eq:treepotexpand}
V_\text{eff} = V + \left(\partial_\rho V \delta \rho + \frac12 \partial^2_\rho V \delta \rho^2 + \text{c.c.} \right) + \partial_\rho \partial_{\bar \rho} V \delta \rho \delta \bar \rho + \mathcal O(\delta \rho^3)\,,
\end{align}
and eliminate $\delta \rho$ demanding that $V_\text{eff}$ be minimized, i.e., $\partial_{\delta \rho}V_\text{eff} = \partial_{\delta \bar \rho} V_\text{eff} = 0$. At second order in $S$ this yields
\begin{align}\label{eq:Ftermeffpot}
V_\text{eff} = \frac{\lambda^2 v^4}{8 \sigma_0^3} \left[2 \sigma_0+ \chi \left(S^2 + \bar S^2 \right) -\frac{\chi^2 + 2}{3} |S|^2 \right] + \mathcal O(|S|^3)\,.
\end{align}
Evidently, there are two possible values of $\chi$ which allow for a vanishing mass of $\text{Re}\,S$ and $\text{Im}\, S$, respectively, and hence for flat directions suitable for inflation,
\begin{align}\label{eq:chivalues}
\chi = \pm(3 \pm \sqrt7)\,.
\end{align}
However, it turns out that for any value of $\chi$, either $\text{Re}\,S $ or $\text{Im}\,S$ has a tachyonic mass, since
\begin{subequations}
\begin{align}
m^2_{\text{Re}\,S} &= -\frac{\lambda^2 v^4}{12 \sigma_0^2} \left(\chi^2 - 6 \chi + 2 \right)\,,\\
m^2_{\text{Im}\,S} &= -\frac{\lambda^2 v^4}{12 \sigma_0^2} \left(\chi^2 + 6 \chi + 2 \right)\,.
\end{align}
\end{subequations}
Thus, any possible inflationary trajectory is destabilized. Note that tachyonic masses of this order cannot be canceled by masses stemming from the Coleman--Weinberg one-loop potential.
Therefore, minimal F-term hybrid inflation appears impossible in this simple setup of moduli stabilization. This conclusion leads us to consider a model of D-term hybrid inflation, where the moduli corrections to the inflationary sector are negligible.

%
%%
%%%
%%%%
%%%%%%%%%%%%%%%%%%%%%%%%%%%%%%%%%%%%%%%%%%%%%%%
%%%%%%%%%%%%%%%%%%%%%%%%%%%%%%%%%%%%%%%%%%%%%%%
%%%%%%%%%%%%%%				 %%%%%%%%%%%%%%%%%%%%%%%
%%%%%%%%%%%%%	%         Section 3        %%%%%%%%%%%%%%%%%%%%%%%
%%%%%%%%%%%%%%				  %%%%%%%%%%%%%%%%%%%%%%
%%%%%%%%%%%%%%%%%%%%%%%%%%%%%%%%%%%%%%%%%%%%%%%
%%%%%%%%%%%%%%%%%%%%%%%%%%%%%%%%%%%%%%%%%%%%%%%

\section{D-term hybrid inflation} 
\label{sec:3}

In D-term inflation the picture is quite different from the previously discussed case. It has the appealing feature that a GUT-scale Fayet--Iliopoulos term\footnote{The consistency of a constant FI-term in supergravity is a subtle issue~\cite{Binetruy:2004hh, Komargodski:2010rb, Dienes:2009td}, which we do not address in this paper. In this context, an interesting approach was used in \cite{Brax:2006yq}, generating an effective FI-term from vacuum expectation values in the modulus sector.} can be naturally generated from anomalous $U(1)$ symmetries in certain string compactifications \cite{Dine:1987xk,Binetruy:2004hh}. This FI-term, together with quantum corrections to the scalar potential, drives inflation. Although D-term inflation is well motivated from string theory, it is necessary to check wether a consistent stabilization of all moduli is possible\footnote{Note that the coupling to a KKLT-type modulus sector using a different K\"ahler potential has been investigated in \cite{Brax:2006yq} along similar lines. For a recent discussion and further references, see \cite{Hebecker:2012aw}.}.

The superpotential of D-term hybrid inflation reads
\begin{align}\label{eq:Dhybsuperpot}
W_\text{DI} = \lambda S \phi_+ \phi_- \,.
\end{align}
In pure D-term inflation without moduli stabilization, using a no-scale K\"ahler potential for the relevant fields results in an F-term potential equivalent to the one of F-term hybrid inflation with $v=0$. The inflationary trajectory corresponds to a flat direction along $\phi_\pm=0$. The D-term potential is generated by the FI-term $\xi$ and the waterfall fields which have non-zero charges under a $U(1)$ gauge symmetry with coupling $g$. During inflation, it induces a vacuum energy $V_0 = \frac{g^2 \xi^2}{2}$. For a detailed description of D-term inflation with canonical K\"ahler potential, see \cite{Binetruy:1996xj,Halyo:1996pp}.

%
%%
%%%
%%%%
%%%%%%%%%%%%%%%%%%%%%%%%%%%%%%%%%%%%%%%%%%%%%%%
%%%%%%%%%%%%%%%%%%%%%%%%%%%%%%%%%%%%%%%%%%%%%%%
\subsection{Moduli corrections \label{subsec_moduli_corrections}}

In our model the superpotential is given by
\begin{align}\label{eq:Dcombinedsuperpot}
W = W_\text{KL} + W_\text{DI}\,,
\end{align}
and the K\"ahler potential is the same as in eq.~\eqref{eq:Ftermkahler}. To determine the influence of the modulus sector on the inflation sector we proceed as in the F-term case, i.e., we expand the potential in the displacement $\delta \rho$, minimize it, and investigate the resulting effective potential for $S$ and $\phi_\pm$. Before integrating out the modulus, the scalar potential is given by $V = V_\text{F} + V_\text{D}$, with
\begin{subequations}\label{eq:Dtermtreepot}
\begin{align}\nonumber\label{eq:Dtermtreepot1}
V_\text{F} &= \frac{1}{X^2} \Bigg\{ \lambda^2 |S|^2 \left( |\phi_+|^2 + |\phi_-|^2 \right)+\lambda^2 |\phi_+ \phi_-|^2  \\ \nonumber
 &\hspace{1.3cm} + \frac13 \left[ \rho + \bar \rho+ \frac{\chi}{6} \left( S^2+\bar S^2 \right) +\frac13 \chi^2 |S|^2 \right] |W'_\text{KL}|^2 \\ 
 &\hspace{1.3cm} - \left[ \left( \overline W_\text{KL} - \frac{\chi}{3} \lambda S \bar \phi_+ \bar \phi_- \right) W'_\text{KL} + \text{c.c.} \right] \Bigg\} \,, \\ \label{eq:Dtermtreepot2}
 V_\text{D} &= \frac{g^2}{2} \left[ \frac{q}{X} \left( |\phi_+|^2 - |\phi_-|^2 \right) - \xi \right]^2 \,,
\end{align}
\end{subequations}
with $X$ as defined in eq.~\eqref{eq:Ftermkahler}. Since $V_\text{eff}$ is much more complicated than the compact expression in the F-term scenario, cf.~eq.~\eqref{eq:treepotexpand}, we restrict ourselves to providing the moduli corrections to the most important quantities. These are, in particular, the scalar masses in the inflation sector. 

\medskip

The inflaton receives a non-zero mass contribution not only from the non-vanishing derivative of $W_\text{KL}$ in eq.~\eqref{eq:Dtermtreepot1}, but also from terms which arise after performing the expansion eq.~\eqref{eq:treepotexpand}, i.e., from integrating out the modulus. However, the resulting correction is zero to first order in $W_\text{KL}$ and $W_\text{KL}'$ and can thus be neglected since $W_\text{KL}\text{, }W'_\text{KL} < \mathcal O (10^{-6})$ for typical values of the racetrack parameters, which renders the corrections much smaller than the contributions from the Coleman--Weinberg potential. Remember that $W_\text{KL}$ and its derivative have to be evaluated at values of $\rho$ slightly shifted from $\sigma_0$, thus yielding non-zero results. The same order of suppression applies to the correction of the first derivative of the scalar potential, proportional to the slow-roll parameter $\epsilon$. This justifies treating $S$ as a flat direction of the tree-level scalar potential of the combined theory, as in the pure D-term case.

\medskip

Corrections to the masses of the waterfall fields are small as well. The end of inflation occurs when one of the waterfall fields obtains a tachyonic mass. Thus, large corrections to the waterfall masses can have grave consequences for the inflationary dynamics. Following the same procedure as for the inflaton mass, we obtain
\begin{align}
m_{\phi_\pm}^2=m_{\phi_\pm,0}^2 + \Delta m_{\phi_\pm}^2 \! \! \left(W_\text{KL},W'_\text{KL}, ... \right)\,,
\end{align}
where
\begin{align}\label{eq:WFuncorrected}
m_{\phi_\pm,0}^2 = \frac{\lambda^2 |S|^2}{X_0} \mp g^2 q \xi\,.
\end{align}
The latter, with $X_0 = 2 \sigma_0 - \frac13 |S|^2 - \frac{\chi}{6}(S^2 + \bar S^2)$ after integrating out the modulus, is the standard result from pure D-term inflation. The leading order corrections are of the form 
\begin{align}\label{eq:WFcorrections}
 \Delta m_{\phi_\pm}^2 = \frac{2 Y_0 W_\text{KL}'-6W_\text{KL}}{Y_0 X_0 W_\text{KL}''} \left( \frac{2 \lambda^2 |S|^2}{X_0} \mp g^2 q \xi \right)
+\mathcal O \left( {W_\text{KL}}^2,{W_\text{KL}'}^2,W_\text{KL} W_\text{KL}',... \right) \,,
\end{align}
with $Y_0= X_0 + \frac13 |S+\chi \bar S|^2$. Note that these corrections are parametrically larger than the ones found in \cite{Brax:2006yq}, due to effective mass terms stemming from the expansion in $\delta \rho$. However, since $W_\text{KL},W'_\text{KL} \ll W''_\text{KL} \sim m_\rho^2$, the correction $\Delta m^2_{\phi_\pm}$ is still negligibly small and does not influence the dynamics of inflation significantly. Moreover, there are no corrections which cause $\phi_\pm$ to be stabilized away from the origin.

%
%%
%%%
%%%%
%%%%%%%%%%%%%%%%%%%%%%%%%%%%%%%%%%%%%%%%%%%%%%%
%%%%%%%%%%%%%%%%%%%%%%%%%%%%%%%%%%%%%%%%%%%%%%%
\subsection{Superconformal symmetry and the Starobinsky model} 
\label{sec:Starobinsky}

Having identified a promising D-term hybrid inflation model with stabilized moduli, we now turn to the phenomenological consequences of this model. Interestingly, during inflation this model is actually equivalent to a model based on a superconformal symmetry \cite{Buchmuller:2012ex}. There, the superpotential is identical to the one in eq.~\eqref{eq:Dhybsuperpot} and the K\"ahler potential reads
\begin{align}\label{eq:SCKahler}
K_\text{SC} = -3 \ln{\left( -\frac13 \Phi \right)}\,,
\end{align}
where 
\begin{align}
\Phi = -3 + |\phi_+|^2 + |\phi_-|^2 + |S|^2 + \frac{\chi}{2} \left( S^2 + \bar S^2 \right)\,,
\label{eq_frame_function}
\end{align}
is the so-called frame function. This type of frame function characterizes a large class of models, dubbed canonical superconformal supergravity models in \cite{Ferrara:2010in}. They feature a remarkably simple structure in the Jordan frame with canonical kinetic terms and a scalar potential which closely resembles that of global supersymmetry. The superconformal symmetry, which is the starting point in constructing these models, is explicitly broken by gauge fixing the so-called compensator field, resulting in the appearance of the Planck scale and the FI-term, and by the term proportional to $\chi$ in eq.~\eqref{eq_frame_function}. This particular symmetry breaking structure allows to keep the attractive features implied by the superconformal symmetry, cf.\ \cite{Ferrara:2010in, Buchmuller:2012ex} for details.

In \cite{Buchmuller:2012ex} the D-term scalar potential is found to be
\begin{align}
V_\text{D} = \frac{g^2}{2} \left[ \Omega^2 q \left(|\phi_+|^2 - |\phi_-|^2 \right) - \xi \right]^2\,,
\end{align}
with $\Omega^2 = -\frac{3}{\Phi}$. It is straightforward to verify that this is identical to eq.~\eqref{eq:Dtermtreepot2} after rescaling
\begin{align}\label{eq:SCrescaling}
S = \sqrt{\rho + \bar \rho} \, S' \,,\qquad \phi_\pm = \sqrt{\rho + \bar \rho} \, \phi_\pm'\,.
\end{align}
The F-term scalar potential is determined by the K\"ahler function $K + \ln |W|^2$, which is invariant under the transformation~\eqref{eq:SCrescaling} since
\begin{align}
K_\text{SC}(S,\phi_\pm) &= -3 \ln{\left(\sqrt{\rho + \bar \rho}\right)}+ \ln{\Omega^{-2}(S',\phi_\pm')}\,, \nonumber \\
\ln{|W(S,\phi_\pm)|^2} &= +3 \ln{\left(\sqrt{\rho + \bar \rho}\right)}+\ln{|W(S',\phi_\pm')|^2}\,. 
\end{align}
Hence, even after rescaling the discussion of Section~\ref{subsec_moduli_corrections} remains valid and the F-term potential vanishes along the inflationary trajectory, as it does in in the model of \cite{Buchmuller:2012ex}. 

Along the inflationary trajectory, the two models thus feature the same scalar potential, allowing us to apply the analysis of \cite{Buchmuller:2012ex} to the model presented here. Here we merely summarize the most important results: We find a two-field inflation model with an attractor solution along the real (imaginary) axis for negative (positive) values of $\chi$. At the end of hybrid inflation, cosmic strings are formed. The spectral index can be as low as $n_s \approx 0.96$. However, for generic values of the gauge coupling $g$ and the $U(1)$ charges $\pm q$ of the waterfall fields, this leads to a too large cosmic string tension, violating the bound obtained from the recent Planck results~\cite{Ade:2013xla}. 

This problem can be circumvented by choosing a relatively large value for $g q$, i.e.,\linebreak $10{ \gtrsim g q \gtrsim \frac{10}{|\chi|}}$, cf.\ \cite{Buchmuller:2013zfa}. In this case agreement with all Planck results can be achieved, including the cosmic string bound~\cite{Ade:2013xla,Ade:2013uln}. Remarkably, in the large-field regime and for an inflationary trajectory along the attractor solution, the model is asymptotically equivalent to the Starobinsky model \cite{Starobinsky:1980te}. In particular, to leading order in $1/N_*$, with $N_*$ the number of e-folds elapsed after the reference scale of the CMB fluctuations exited the horizon, the scalar spectral index, the tensor-to-scalar ratio, and the running of the spectral index are given by
\begin{equation}
 n_s \approx 1- \frac{2}{N_*} \,, \qquad r \approx \frac{12}{N_*} \,, \qquad \frac{\text{d}n_s}{\text{d} \ln \! k} \approx - \frac{2}{N_*^2} \,,
\end{equation}
which, for $N_* \approx 55$, describes the Planck data very well \cite{Ade:2013uln}.\footnote{Note that inflation terminates at $S_\eta \approx 1$, cf.~\cite{Buchmuller:2013zfa}, so that corrections to the K\"ahler potential eq.~\eqref{eq:Ftermkahler}, suppressed by inverse powers of the Planck mass, may be relevant.} For $g^2 \approx \frac12$, as expected for a GUT gauge coupling, requiring the correct normalization of the scalar contribution to the primordial fluctuations fixes the FI-term at roughly the GUT scale, $\sqrt{\xi} \approx 7.7 \times 10^{15}\,\text{GeV}$. For example, for $q = 8$ this implies a cosmic string tension of $G \mu \approx 3.16 \times 10^{-7}$, very close the recent Planck limit $G \mu < 3.2 \times 10^{-7}$~\cite{Ade:2013xla}. This large value of $q$ is problematic from the point of view of GUT model building, which suggests to explore alternative ways to satisfy the cosmic string bound, cf.~\cite{Buchmuller:2012ex}.

%
%%
%%%
%%%%
%%%%%%%%%%%%%%%%%%%%%%%%%%%%%%%%%%%%%%%%%%%%%%%
%%%%%%%%%%%%%%%%%%%%%%%%%%%%%%%%%%%%%%%%%%%%%%%
\subsection{Low-energy supersymmetry breaking}

During inflation the D-term inflation model under consideration exhibits a positive vacuum energy $V_0 = \frac{g^2 \xi^2}{2}$ and thus, supersymmetry is broken. After inflation has ended, however, one of the waterfall fields receives a vacuum expectation value which causes $V_\text{D}$ to vanish identically, while the other one and the inflaton are stabilized at the origin. It then follows that $V_\text{F} = V_\text{D} = m_{3/2} = 0$ after inflation, i.e., supersymmetry is restored. In view of low-energy phenomenology, it is thus necessary to check wether the presented model can be combined with a separate sector of supersymmetry breaking without spoiling either inflation or moduli stabilization.

A  simple way of breaking supersymmetry is adding a quantum corrected O'Raifeartaigh model with the following K\"ahler potential and superpotential for a chiral `Polonyi' field $P$ \cite{Kallosh:2006dv},
\begin{align}\label{eq:Polonyisector}
K_\text{P} = |P|^2 - \frac{|P|^4}{\Lambda^2}\,, \qquad W_\text{P} = \mu^2 P\,.
\end{align}
Here, a heavy field of mass $\Lambda \ll 1$ has been integrated out, and $\mu^2$ is the scale of supersymmetry breaking. In addition, to allow for a small or vanishing cosmological constant we tune the value of $W_0$ away from the KL-value eq.~\eqref{eq:KLW0} by an amount $\Delta W_0$. In an underlying string compactification this is achieved by slightly tuning the flux quanta which determine the vacuum expectation value of the Gukov--Vafa--Witten potential. As a result, a complete model with broken supersymmetry can be defined by
\begin{align}\label{eq:polonyicombined}
K = -3 \ln{X} + K_\text{P} \,, \qquad W = W_\text{KL} + W_\text{DI} + W_\text{P} + \Delta W_0\,.
\end{align}
Note that the supersymmetry breaking sector is not of no-scale form. This is phenomenologically required for low-energy supersymmetry breaking \cite{Dudas:2012wi}. The derivation of this K\"ahler potential from a higher-dimensional theory remains an open problem.

The compatibility of this supersymmetry breaking mechanism with moduli stabilization has been studied in \cite{Kallosh:2006dv,Linde:2011ja,Dudas:2012wi}. The constant $\Delta W_0$ shifts the Minkowski minimum of the potential to an AdS minimum with $V_\text{AdS} \approx -\frac{3 (\Delta W_0)^2}{8 \,\sigma_0^3}$ at roughly the same value of $\sigma_0$. The uplift due to the Polonyi field raises the value of $V$ in the minimum to zero if
\begin{align}\label{eq:Minkcondition}
\Delta W_0 = \frac{\mu^2}{\sqrt3}\,,
\end{align}
resulting in a Minkowski vacuum with broken supersymmetry. In this vacuum the gravitino mass is given by
\begin{align}\label{eq:m32}
m_{3/2}^2 \approx \frac{\mu^4}{24 \,\sigma_0^3}\,,
\end{align}
at leading order in $\mu^2$ and $\Lambda$.

In this minimum the Polonyi field is stabilized on the real axis at $P_0 \approx \frac{\sqrt 3}{6} \Lambda^2$. Moreover, it is possible to decouple the Polonyi field before the beginning of inflation, i.e., at masses larger than the inflationary Hubble scale. We can achieve a mass hierarchy 
\begin{align}\label{eq:Polhierarchy}
m_\rho > m_P > H_\text{inf} \gg m_{3/2}\,,
\end{align}
by appropriately choosing $\mu$, $\Lambda$, and the parameters in $W_\text{KL}$. Specifically, $m_P^2$ in the Minkowski minimum reads
\begin{align}\label{eq:Polonyimass}
m_P^2 \approx \frac{\mu^4}{2 \sigma_0^3 \Lambda^2} \gg m_{3/2}^2\,.
\end{align}
Notice that we have used $\sigma_0$ in all of the above expressions because the back-reaction of the shift $\delta \rho$ on the dynamics of the Polonyi field is negligible. However, it is important to keep in mind that $\rho$ is still slightly shifted away from $\sigma_0$ due to the presence of $P$ and $\Delta W_0$, so that $W_\text{KL}, W_\text{KL}'  \neq 0$. Requiring that $m_P \gtrsim H_\text{inf}$ and demanding $\Lambda \gtrsim \mu$ in the effective theory~\eqref{eq:Polonyisector} leads to the lower bounds $\mu, \, \Lambda \gtrsim 10^{-5}$ for typical values of the racetrack parameters\footnote{Here we have used $H_\text{inf} \sim 0.1 M_\text{GUT}^2$, with $M_\text{GUT} \gtrsim 10^{-3}$.}. 

Remarkably, as a consequence GUT-scale inflation implies a stringent lower bound on the gravitino mass. From eqs.~\eqref{eq:m32}\,-\,\eqref{eq:Polonyimass} one obtains
\begin{align}\label{eq:m32bound}
m_{3/2}^2 \gtrsim 0.1 \Lambda^2 \,H_\text{inf}^2 \gtrsim 10^{-25}\,.
\end{align}
Starting from the KL model for moduli stabilization, one may have expected that an arbitrarily small value of the gravitino mass is possible. However, since both $m_P$ and the mass scale $\Lambda$ are constrained by the GUT scale, one is driven to a regime of `high-scale supersymmetry' with $m_{3/2} \gtrsim 10^5 \, \text{GeV}$. Even if the Polonyi field is allowed to be lighter than $H_\text{inf}$ but heavier than the inflaton, thus taking part in the dynamics of inflation, this bound is not significantly relaxed.

Notice that the choice of parameters in the Polonyi sector only slightly influences the modulus sector and vice versa. Therefore, in a large portion of parameter space the proposed mechanism of supersymmetry breaking does not interfere with moduli stabilization. Especially, even if $\mu$ is chosen to be very large compared to the GUT scale, additional tuning of $\Delta W_0$ will always prevent destabilization of the modulus.

\medskip

Quantifying the impact of the Polonyi field on the inflationary dynamics is slightly more involved. As in our previous discussion of moduli corrections to the inflaton sector, the impact on $\epsilon$, the inflaton mass, and the waterfall masses has to be evaluated. In order to consider all possible terms, we proceed along the lines of Section~\ref{subsec_moduli_corrections} and take a possible shift $\delta P$ during inflation into account, as well as corrections resulting from integrating out the modulus. This results in the following corrections:
\begin{itemize}

\item The mass of the inflaton, which can be chosen to be the real part of $S$, receives the correction
\begin{align}\label{eq:Polcorrectioninfmass}
\Delta m_{\text{Re}\,S}^2 = m_{3/2}^2 \left(1+ \chi \right)^2\,,
\end{align}
at leading order in $\mu^2$ and $S^2$.\footnote{We have assumed that $2\sigma_0 \gg |S|^2$ towards the end of inflation, which is satisfied even in the large field regime discussed in \cite{Buchmuller:2013zfa}.} Note that this term is present even before integrating out $\rho$, which yields small corrections of higher order in $S^2$. Eq.~\eqref{eq:Polcorrectioninfmass} implies that successful inflation also puts an upper bound on the gravitino mass\footnote{We thank the referee for pointing this out.}, unless $\chi =-1$, which corresponds to a shift-symmetric K\"ahler potential. For $\chi \neq -1$, demanding that the correction to the slow-roll parameter $\eta$ does not alter the prediction for $n_s$ by more than $1\sigma$, cf.~the recent Planck data \cite{Ade:2013uln}, leads to $m_{3/2} \lesssim 10^{10} \, \text{GeV}/|1+\chi|$. The bound resulting from the correction to the slow-roll parameter $\epsilon$ is less severe.

\item The leading order mass correction to the waterfall fields originates solely from the effective potential $V_\text{eff}$ where the modulus has been integrated out, analog to the corrections in eq.~\eqref{eq:WFcorrections}. Specifically,
\begin{align}\label{eq:Polcorrectionwfmass}
\Delta m_{\phi_\pm}^2 = \frac{\mu^2 \left( S^2 + \bar S^2 + 2 \chi |S|^2 \right)}{16 \sqrt3  \sigma_0^3 W_\text{KL}''}\cdot m_{\phi_\pm,0}^2\,,
\end{align}
with $m_{\phi_\pm,0}$ defined by eq.~\eqref{eq:WFuncorrected}. Depending on the size of $\mu$ these corrections can be parametrically larger than the ones from the modulus sector, cf.~eq.~\eqref{eq:WFcorrections}. However, since $\mu^2$ is smaller than $W_\text{KL}''$ for typical racetrack parameter values, $\Delta m_{\phi_\pm}^2$ is still suppressed by at least three orders of magnitude compared to $m_{\phi_\pm,0}^2$.

\end{itemize}

\medskip

We conclude that our model can be extended by a simple supersymmetry breaking sector without spoiling any of its features. In this setup, the gravitino mass has to satisfy lower and upper bounds,\begin{align}\label{eq:m32bounds}
10^{5}\, \text{GeV} \lesssim m_{3/2} \lesssim 10^{10}\, \text{GeV}\,,
\end{align}
which are due to the high scale of inflation and the slow roll conditions, respectively.
%
%%
%%%
%%%%
%%%%%%%%%%%%%%%%%%%%%%%%%%%%%%%%%%%%%%%%%%%%%%%
%%%%%%%%%%%%%%%%%%%%%%%%%%%%%%%%%%%%%%%%%%%%%%%
%%%%%%%%%%%%%%				 %%%%%%%%%%%%%%%%%%%%%%%
%%%%%%%%%%%%%	%         Section 4        %%%%%%%%%%%%%%%%%%%%%%%
%%%%%%%%%%%%%%				  %%%%%%%%%%%%%%%%%%%%%%
%%%%%%%%%%%%%%%%%%%%%%%%%%%%%%%%%%%%%%%%%%%%%%%
%%%%%%%%%%%%%%%%%%%%%%%%%%%%%%%%%%%%%%%%%%%%%%%

\section{Conclusion} 
\label{sec:Conclusion}

In light of the recent Planck data, slow-roll inflation remains a very successful paradigm for the earliest stages of our universe. Realizing this paradigm in a concrete UV-completed particle physics theory, however, faces a number of challenges, including the identification of the particle physics nature of the inflaton, a possible embedding in string theory and the connection to supersymmetry breaking after inflation. Here, we propose a model of supersymmetric hybrid inflation which allows for racetrack moduli stabilization, as employed in certain type IIB string compactifications, as well as for supersymmetry breaking by means of a quantum corrected O'Raifeartaigh model, while simultaneously explaining the cosmological parameters measured by the Planck satellite.

Using the standard no-scale K\"ahler potential, augmented by a symmetry breaking term, we find that F-term hybrid inflation is unfeasible. Generically, the inflaton mass receives large corrections, spoiling slow-roll inflation. While this can be remedied by tuning the symmetry breaking parameter $\chi$, the presence of a large tachyonic mass destabilizing any potential inflationary trajectory is unavoidable. However, supersymmetric D-term hybrid inflation is not plagued by this problem. Tracking the evolution of the modulus field during inflation and integrating out the modulus, we find that the corrections to the inflationary dynamics induced by the modulus sector are small. If the modulus is stabilized before the onset of inflation, i.e., $m_\rho > H_\text{inf}$, we obtain an effective inflation model which, along the inflationary trajectory, is identical to superconformal D-term inflation.

Concerning the inflationary predictions, i.e., amplitude and spectral indices of the CMB power spectrum, we find very good agreement with the recent Planck data. Generically, cosmic strings produced at the end of D-term inflation exhibit a string tension exceeding current bounds. However, viable regions of parameter space remain, for large values of the waterfall $U(1)$ charge $q$. In the large-field regime the scalar potential of the inflaton field is identical to that of the Starobinsky model.

In order to account for supersymmetry breaking in the Minkowski vacuum after inflation, we add a quantum corrected O'Raifeartaigh model. We calculate possible interactions between the inflation, modulus, and Polonyi field sector. We find that the only displacement of the modulus minimum resulting in relevant corrections is the one stemming from the slow-roll of the inflaton. Generically, however, all these corrections turn out to be small, allowing for an effectively decoupled supersymmetry breaking sector. The gravitino mass is constrained to the range $10^{5}\, \text{GeV} \lesssim m_{3/2} \lesssim 10^{10}\, \text{GeV}$.

\medskip

In summary, we present a working model of inflation, successfully combined with KL moduli stabilization and supersymmetry breaking and in accordance with experimental data. Further interesting questions concern the embedding of our model into a higher-dimensional GUT or string model, and the implications for low-energy particle phenomenology. 
%
%
%
%%
%%%
%%%%
%%%%%%%%%%%%%%%%%%%%%%%%%%%%%%%%%%%%%%%%%%%%%%%
%%%%%%%%%%%%%%%%%%%%%%%%%%%%%%%%%%%%%%%%%%%%%%%
\subsection*{Acknowledgments}
The authors thank Ido Ben-Dayan and Alexander Westphal for helpful discussions. This work has been supported by the German Science Foundation (DFG) within the Collaborative Research Center 676 ``Particles, Strings and the Early Universe''. The work of C.W. is supported by a scholarship of the Joachim Herz Stiftung.

%
%%
%%%
%%%%
%%%%%%%%%%%%%%%%%%%%%%%%%%%%%%%%%%%%%%%%%%%%%%%
%%%%%%%%%%%%%%%%%%%%%%%%%%%%%%%%%%%%%%%%%%%%%%%
%%%%%%%%%%%%%%				 %%%%%%%%%%%%%%%%%%%%%%%
%%%%%%%%%%%%%	%         Bibliography   %%%%%%%%%%%%%%%%%%%%%%%
%%%%%%%%%%%%%%				  %%%%%%%%%%%%%%%%%%%%%%%
%%%%%%%%%%%%%%%%%%%%%%%%%%%%%%%%%%%%%%%%%%%%%%%
%%%%%%%%%%%%%%%%%%%%%%%%%%%%%%%%%%%%%%%%%%%%%%%

%\bibliographystyle{utphys}

\end{document}